\documentclass[english,prl,twocolumn,groupedaddress,reprint]{revtex4}
\usepackage{amsfonts}
\usepackage[T1]{fontenc}
\usepackage[latin9]{inputenc}
\usepackage{color}
\usepackage{amsmath}
\usepackage{amssymb}
\usepackage{graphicx}
\usepackage{esint}
\usepackage{comment}
\usepackage{bbold}
\usepackage{tikz}
\usepackage{babel}
\usepackage{minitoc}
\usepackage{bm}

\begin{document}

\title{Amplification effects in optomechanics via weak measurement}
\author{Gang Li\footnote{ligang0311@sina.cn},$^{1}$ Tao Wang,$^{2}$ and He-Shan Song\footnote{
hssong@dlut.edu.cn}$^{1}$}
\affiliation{$^{1}$School of Physics and Optoelectronic Technology, Dalian University of
Technology, Dalian 116024, People's Republic of China\\
$^{2}$College of Physics, Jilin University, Changchun 130012, People's Republic of China}
\date{\today }

\begin{abstract}
We revisit the scheme of single-photon weak-coupling optomechanics using post-selection, proposed by Pepper, Ghobadi, Jeffrey, Simon and Bouwmeester [Phys. Rev. Lett. \textbf{109}, 023601 (2012)], by analyzing the exact solution of the dynamical evolution. Positive and negative amplification effects of the displacement of the mirror's position can be generated when the \emph{Kerr phase} is considered. This effect occurs when the post-selected state of the photon is orthogonal to the initial state, which can not be explained by the usual weak measurement results. The amplification effect can be further modulated by a phase shifter, and the maximal displacement state can appear within a short evolution time.
~~~~~\newline~~~~~\newline
\end{abstract}

\maketitle

\section{I. INTRODUCTION}

Post-selected weak measurement, first discovered by Aharonov, Albert and Vaidman \cite{Aharonov88}, has been a subject full of curiosity and can be used to explain some counter-intuitive quantum paradoxes \cite{Aharonov05}. Its important characteristic is found useful to measure and amplify small physical quantities or effects, and has applications in direct measurement of wave function \cite{Bamber11}, spin hall effect of light \cite{Hosten08}
and ultrasensitive beam deflection \cite{Howell09}. Although weak measurement has many applications, its application in quantum optomechanics is seldom reported. Quantum optomechanical system usually refers to a high finesse cavity with a movable mirror where the light in the cavity can give a force on the mirror \cite{Girvin09,Marquardt13}. When there is only one photon in the cavity, the displacement of the mirror caused by the photon is
hard to be detected since it is much smaller than the spread of the mirror wave packet. In a recent paper \cite{Bouwmeester12}, Pepper et al proposed a scheme to generate the one phonon state of the mirror by using a single photon in a special quantum optomechanical system, where a weak measurement is implemented through a nested Mach-Zehnder interferometer. Their result
motivates a possibility to detect the displacement of the mirror caused by one photon using weak measurement.

In the scheme proposed by Pepper et al, a \emph{Kerr phase} \cite{Mancini97,Bose97} proportional to $(a^{\dag }a)^{2}$ was omitted. In this paper, we will revisit their scheme but retain the \emph{Kerr phase}. When combined with a post-selected weak measurement, it will be shown that the \emph{Kerr phase} can lead to an amplification effect of the mirror's
displacement. And the maximal amplification value can reach the level of the ground state fluctuation and is detectable in principle. Even more surprisingly, besides the positive amplification effect, there is a counter-intuitive negative amplification. Furthermore it is very easy for the amplification effect to be modulated by a phase shifter. Our result also
shows that some special states of the mirror, including the states achieving the maximal displacement, i.e., the equal superposition of the ground state and the one photon state, can appear within a short evolution time, which is in sharp contrast to the result in \cite{Bouwmeester12} where the tiny \emph{Kerr phase} was omitted and only the one photon state is achieved. We note that our scheme uses Gaussian state as the initial state of the pointer as
in Ref. \cite{Aharonov88,Pang12}, but we get the amplification effect when the post-selected state is chosen to be orthogonal to the initial state of the system.

The structure of our paper is as follows. In Secs. II, we briefly review the post-selected weak measurement. In Secs. III, we state the main result of this work, including the positive amplification and negative amplification effects of the mirror's displacement which is induced by the \emph{Kerr phase}. In Secs. IV, we show that the amplification effect can be modulated by a phase shifter. In Secs. V and VI, we give the discussion and conclusion about the work, respectively.

\section{II. POST-SELECTED WEAK MEASUREMENT}

Firstly we simply review the post-selected weak measurement \cite{Aharonov88}. In the interaction picture, suppose the Hamiltonian of the quantum system to be measured and the quantum pointer is
\begin{equation}
\hat{H}=\hbar \chi (t)\hat{\sigma}_{z}\otimes \hat{p},~~~~~\chi (t)=\chi\delta (t-t_{0})
\end{equation}
where $\hbar $ is Planck's constant, $\chi $ is a small coupling constant, $\hat{\sigma _{z}}$ is a spinlike observable acting on the quantum system to be measured and $\hat{p}$ is the momentum operator of the quantum pointer that conjugates to the position operator $\hat{q}$. The time factor $\delta(t-t_{0})$ means the time of weak interaction is a very short instant.
Suppose that the initial states of the system and the pointer are $|\psi_{i}\rangle $ and $|\phi \rangle $, respectively. When the weak interaction is finished, the time evolution of the total system (the system and the pointer) is given by
\begin{equation}
e^{-i\chi \hat{\sigma}_{z}\otimes \hat{p}}|\psi _{i}\rangle \otimes |\phi\rangle.
\end{equation}
When the post-selected state $|\psi _{f}\rangle $ is projected onto the
system, the final state of the pointer is
\begin{equation}
|\phi ^{\prime }\rangle =\frac{\langle \psi _{f}|e^{-i\chi \hat{\sigma}_{z}\otimes \hat{p}}|\psi _{i}\rangle \otimes |\phi \rangle }{|\langle \psi_{f}|e^{-i\chi \hat{\sigma}_{z}\otimes \hat{p}}|\psi _{i}\rangle \otimes|\phi \rangle |},
\end{equation}
which is usually called postselection. The average displacement of the pointer variable $\hat{M}$ after the post-selected weak measurement is given by
\begin{equation}
\langle M\rangle =\langle \phi ^{\prime }|M|\phi ^{\prime }\rangle -\langle\phi |M|\phi \rangle.
\end{equation}
When $|\psi _{f}\rangle $ is not the same as $|\psi _{i}\rangle $, $\langle M\rangle $ may be very large and lead to an amplification effect.

The usual weak measurement theory as well as the induced amplification effect is valid only for a post-selected state of the system not orthogonal to the initial state \cite{Aharonov88,Jozsa07}, i.e., $\langle \psi_{f}|\psi _{i}\rangle \neq 0$. The weak measurement theory with $\langle\psi _{f}|\psi _{i}\rangle =0$ has also been discussed in Ref. \cite{Pang12}, but no amplification effect is found. In this paper we will discuss a weak
measurement with $\langle \psi _{f}|\psi _{i}\rangle =0$, but the amplification effect is obtained.

\section{III. AMPLIFICATION In OPTOMECHANICS}

\subsection{1. The optomechanics model}

\begin{figure}[tbp]
\includegraphics[width=1\columnwidth]{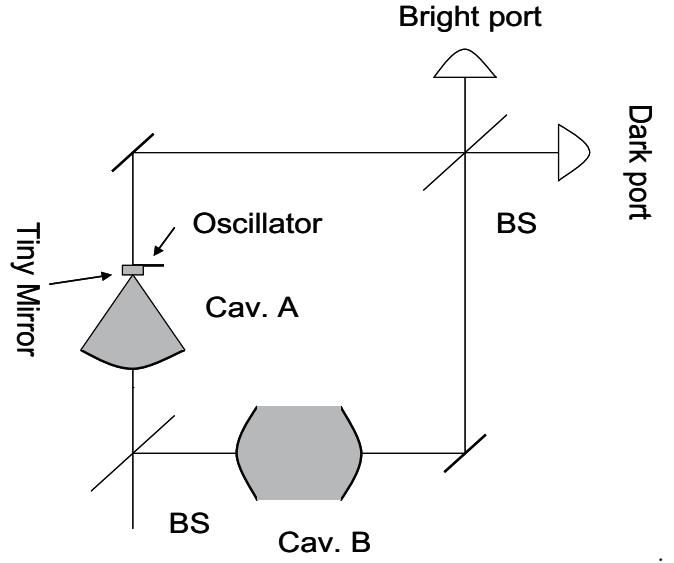}.
\caption{The photon enters the first beam splitter of March-Zehnder interferometer, followed by an optomechanical cavity A and a conventional cavity B. The photon weakly excites the small mirror. After the second beam splitter, dark port is detected, i.e., postselection acts on the case where the mirror has been excited by a photon, and fails otherwise.}
\end{figure}

Now consider a Mach-Zehnder interferometer shown in Fig. 1, which is the same as that considered by Pepper et al. The optomechanical cavity A is embedded in one arm of the March-Zehnder interferometer and a stationary Fabry-P\'{e}rot cavity B is placed in another arm. The two beam splitters are both symmetric. The Hamiltonian of the optomechanical system in the interferometer is expressed as followed:
\begin{equation}
\hat{H}=\hbar \omega _{0}(a^{\dag }a+b^{\dag }b)+\hbar \omega _{m}c^{\dag}c-\hbar ga^{\dag }a(c+c^{\dag }),
\end{equation}
where $\hbar $ is Planck's constant, $\omega _{0}$ is frequency of the optical cavity A, B and the corresponding annihilation operators are $\hat{a}$ and $\hat{b}$, $\omega _{m}$ is frequency of mechanical system and the corresponding annihilation operator is $\hat{c}$, and the optomechanical coupling strength $g=\frac{\omega _{0}}{L}\sigma $, where $L$ is the length
of the cavity A or B, $\sigma =(\hbar /2m\omega _{m})^{1/2}$ which is the zero-point fluctuation and $m$ is the mass of mechanical system. Here $a^{\dag}a$ of the Eq. (5) corresponds to $\hat{\sigma}_{z}$ of the Hamiltonian in the standard scenario of weak measurement and $c+c^{\dag}$ corresponds to $\hat{p}$. It is a weak measurement model where the mirror is used as the pointer to measure the number of photon in cavity A.

According to the results of Ref. \cite{Mancini97,Bose97}, the time evolution operator corresponding to the Hamiltonian of Eq. (5) is given by
\begin{eqnarray}
U(t) &=&\exp [-ir(a^{\dag }a+b^{\dag }b)\omega _{m}t]\exp [i(a^{\dag}a)^{2}\phi (t)] \notag \\
&\times &\exp [a^{\dag }a(\varphi (t)c^{\dag }-{\varphi }^{\ast }(t)c)]\exp[-ic^{\dag }c\omega _{m}t],
\end{eqnarray}
where $\phi (t)=k^{2}(\omega _{m}t-\sin \omega _{m}t)$, $\varphi(t)=k(1-e^{-i\omega _{m}t})$, $r=\omega _{0}/\omega _{m}$, $k=g/\omega _{m}$ is the scaled coupling parameter. We assume that the initial state of the mirror is prepared at the ground state $|0\rangle _{m}$ that can be achieved using sideband-resolved cooling technique \cite{Kippenberg07,Girvin07}. If there is $n$ photon appearing in cavity A, from the time evolution operator it can be found that there exists a phase $e^{in^{2}\phi (t)}$, and the phase will be called \emph{Kerr phase} throughout this paper.

From the literature \cite{Marshall03} we know that if the displacement of the mirror can be detected experimentally it should be not smaller than $\sigma $ , i.e., the zero-point fluctuation of the mirror at the ground state $|0\rangle _{m}$. If there is no cavity B and only one photon in cavity A interacting with the mirror, from the time evolution operator it
can be found that the mirror will be changed from the ground state $|0\rangle _{m}$ to the coherent state $|\varphi (t)\rangle $ (some phase is omitted). The position displacement of the mirror will be $\langle \varphi(t)|\hat{q} |\varphi (t)\rangle $ and it is not more than $4k\sigma $ for any time $t$. Since $k=g/\omega _{m}$ can not be bigger than $0.25$ in
weak coupling condition \cite{Marshall03}, the maximal displacement of the mirror $4k\sigma $ can not be bigger than the zero-point fluctuation of the mirror, therefore the displacement of the mirror caused by one photon can not be detected. In the following we will consider the case the cavity B is added and show how the weak measurement can amplify the mirror's displacement.

\subsection{2. Amplification with post-selected weak measurement in optomechanics}

Suppose a single photon is input into the interferometer shown in Fig. 1, the state of the photon after the first beam splitter becomes
\begin{equation}
|\psi _{i}\rangle =\frac{1}{\sqrt{2}}(|1\rangle _{A}|0\rangle _{B}+|0\rangle_{A}|1\rangle _{B}),
\end{equation}
which is an equal superposition of being in arm $A$ and $B$. The initial state of the mirror is prepared at the ground state $|0\rangle_{m}$. For the completeness of the discussion, the damping of the mirror is also considered \cite{Bose97}, thus the master equation of the optomechanical system is
\begin{eqnarray}
\frac{d\rho(t)}{dt}&=&-\frac{i}{\hbar}[H,\rho(t)]  \notag \\
&+&\frac{\gamma_m}{2}[2c\rho(t)c^{\dag}-c^{\dag}c\rho(t)-\rho(t) c^{\dag}c],
\end{eqnarray}
where $\gamma_{m}$ is the damping constant. After interacting weakly with the optomechanical system, the density matrix of the total system is
\begin{eqnarray}
\rho (t) &=&\frac{1}{2}(|1\rangle _{A}|0\rangle _{B}\langle 1|_{A}\langle0|_{B}\otimes |\varphi (\gamma ,t)\rangle _{m}\langle \varphi (\gamma
,t)|_{m}  \notag \\
&+&e^{i\phi (t)-D(\gamma ,t)}|1\rangle _{A}|0\rangle _{B}\langle0|_{A}\langle 1|_{B}\otimes |\varphi (\gamma ,t)\rangle _{m}\langle 0|_{m}
\notag \\
&+&e^{-i\phi (t)-D(\gamma ,t)}|0\rangle _{A}|1\rangle _{B}\langle1|_{A}\langle 0|_{B}\otimes |0\rangle _{m}\langle \varphi (\gamma ,t)|_{m}
\notag \\
&+&|0\rangle _{A}|1\rangle _{B}\langle 0|_{A}\langle 1|_{B}\otimes |0\rangle_{m}\langle 0|_{m}),
\end{eqnarray}
where $\gamma =\gamma _{m}/\omega _{m}$, $\phi (t)=k^{2}(\omega _{m}t-\sin\omega _{m}t)$, $\varphi (\gamma ,t)=\frac{ik}{i+\gamma /2}(1-e^{-(i+\gamma/2)\omega _{m}t})$ is the amplitude of the coherent states of the mirror generated by one photon and
\begin{eqnarray}
D(\gamma ,t) &=&\frac{k^{2}\gamma }{2(1+\gamma ^{2}/4)}[\omega _{m}t+\frac{1-e^{-\gamma \omega _{m}t}}{\gamma }  \notag \\
&-&\frac{e^{(i-\gamma /2)\omega_{m}t}-1}{i-\gamma/2}+\frac{e^{-(i-\gamma/2)
\omega_{m}t}-1}{i+\gamma /2}].
\end{eqnarray}
The relative phase $\phi(t)$ between the coherent state $|\varphi(\gamma,t)\rangle_{m}$ and the ground state $|0\rangle_{m}$ is the \emph{Kerr phase} in Ref. \cite{Mancini97,Bose97}.

When a photon is detected in the dark port, in the language of weak measurement the post-selected state of the single-photon \cite{Bouwmeester12}is
\begin{equation}
|\psi _{f}\rangle =\frac{1}{\sqrt{2}}(|1\rangle _{A}|0\rangle _{B}-|0\rangle_{A}|1\rangle _{B}),
\end{equation}
which is orthogonal to the initial state, i.e., $\langle \psi _{f}|\psi_{i}\rangle =0$. Then the final state of the mirror becomes
\begin{eqnarray}
\rho _{os}(t) &=&\frac{1}{4}(|\varphi (\gamma ,t)\rangle _{m}\langle \varphi(\gamma ,t)|_{m}-e^{i\phi (t)-D(\gamma ,t)}  \notag \\
&\times &|\varphi (\gamma ,t)\rangle _{m}\langle 0|_{m}-e^{-i\phi(t)-D(\gamma ,t)}  \notag \\
&\times &|0\rangle _{m}\langle \varphi (\gamma ,t)|_{m}+|0\rangle_{m}\langle 0|_{m}).
\end{eqnarray}

\begin{figure}[tbp]
\includegraphics[width=1\columnwidth]{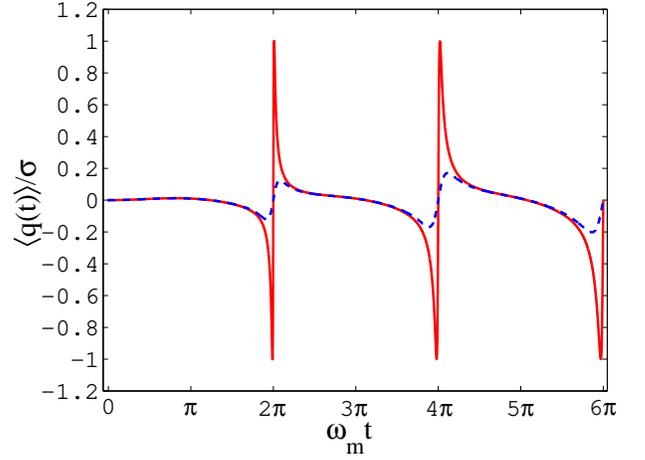}.
\caption{The average displacement $\langle q(t)\rangle /\sigma $ of the mirror as a function of $\omega _{m}t$ with $k=0.005$, $\gamma =0$ (solid line) and $\gamma =0.005$ (dashed line).}
\end{figure}

\begin{figure}[tbp]
\includegraphics[width=1\columnwidth]{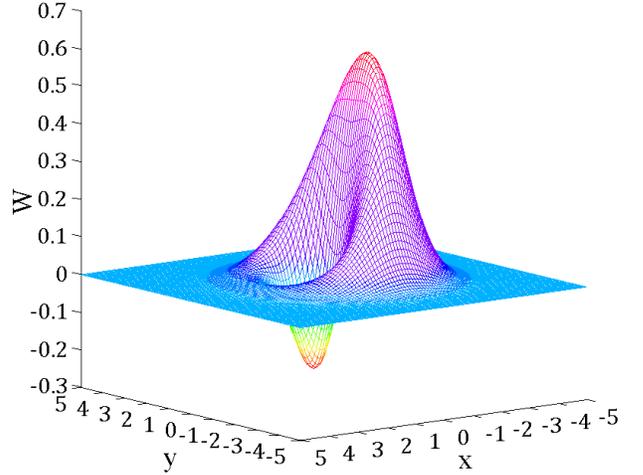}.
\caption{Wigner function of the state achieving the maximal negative amplification. In this figure, $x$ stands for $c+c^{\dag }$ and $y$ for $-i(c-c^{\dag })$. They are therefore dimensionless quantities.}
\end{figure}

The average displacement of the position operator $\hat{q}=(\hat{c}+\hat{c}^{\dag})\sigma$ of the mirror, or the average position of the quantum pointer, is given by
\begin{equation}
\langle q\rangle =\frac{Tr(\rho_{os}\hat{q})}{Tr(\rho_{os})}-Tr(|0\rangle_{m}
\langle0|_{m}\hat{q}).
\end{equation}

After plugging Eq. (12) into Eq. (13) we have
\begin{eqnarray}
\langle q(t)\rangle &=&\sigma \lbrack \varphi (\gamma ,t)+\varphi ^{\ast}(\gamma ,t)-e^{-\frac{|\varphi (\gamma ,t)|^{2}}{2}}  \notag \\
&&(e^{i\phi (t)-D(\gamma ,t)}\varphi (\gamma ,t)+e^{-i\phi (t)-D(\gamma ,t)}
\notag \\
&&\varphi ^{\ast }(\gamma ,t))]/[2-e^{-\frac{|\varphi (\gamma ,t)|^{2}}{2}}(e^{i\phi (t)-D(\gamma ,t)}  \notag \\
&&+e^{-i\phi (t)-D(\gamma ,t)})].
\end{eqnarray}
The average displacement $\langle q(t)\rangle /\sigma $ of the mirror is shown in Fig. 2 as a function of $\omega _{m}t$ with $k=0.005$, $\gamma =0$ (solid line) and $\gamma =0.005$ (dashed line). It can be seen that when the damping is present the two extreme values are both reduced (dashed line), but the actual $\gamma $ can be very small ($\gamma =5\times 10^{-7}$ in
\cite{Bouwmeester12}). The result for $\gamma =5\times 10^{-7}$ is almost the same as the one for $\gamma =0$. The positive and negative amplification effects are very prominent around the vibration periods of the mirror $\omega _{m}t=2n\pi$ $(n=1,2,3,\cdots )$. Both positive and negative amplifications can reach strong coupling limiting (the level of the ground
state fluctuation) \cite{Marshall03} $\langle q\rangle =\pm \sigma $. Note that the maximal displacement of the mirror caused by one photon in cavity A is $4k\sigma $ and the displacement $\pm \sigma $ can be obtained using weak measurement, therefore the amplification factor can be $Q=\pm 1/4k$ which is $\pm 50$ when $k=0.005$. The negative amplification is unusual because
intuitively there can not exist negative displacement relative to the direction of the photon propagation, therefore it is a counter-intuitive result.

\subsection{3. Small quantity expansion about time for amplification}

In order to further explicitly explain the amplification phenomenon displayed in Fig. 2, we take no account of the damping of the mirror, i.e., $\gamma _{m}=0$. Then the state of the Eq. (12) becomes
\begin{equation}
\Psi _{os}(t)=\frac{1}{\sqrt{2}}(e^{i\phi (t)}|\varphi (t)\rangle_{m}-|0\rangle _{m}),
\end{equation}
where $\varphi (t)=k(1-e^{-i\omega _{m}t})$ and $\phi (t)=k^{2}(\omega_{m}t-\sin \omega_{m}t)$. To observe the amplification effects appearing around the vibrational periods of the mirror, we can perform a small quantity expansion about time $T$ till the second order, where $T=2n\pi$ $(n=1,2,\cdots )$. Suppose that $|\omega_{m}t-T|\ll 1$, $k\ll 1$ and $k^{2}T\ll 1$, there is
\begin{equation}
\Psi _{os}(t)\approx ik^{2}T|0\rangle _{m}+ik(\omega _{m}t-T)|1\rangle _{m}.
\end{equation}
We note that in the paper by Pepper et al, the \emph{Kerr phase} $\phi (t)$ is neglected and they find $\Psi _{os}(t)$ proportional to $|1\rangle _{m}$ instead of the above superposition state. Here we retain the \emph{Kerr phase} and use the approximation $e^{i\phi (t)}\approx 1+ik^{2}T$ to get the above superposition state. Substituting Eq. (16) into Eq. (13), the
average value of the displacement operator $\hat{q}$ is given by
\begin{eqnarray}
\langle q(t)\rangle_{|\omega _{m}t-T|\ll1}  &=&2Tk^{3}\sigma (\omega_{m}t-T)/[T^{2}k^{4}  \notag \\
&+&k^{2}(\omega _{m}t-T)^{2}],
\end{eqnarray}
which gets its maximal value $\sigma $ when $Tk^{2}=k(\omega _{m}t-T)$, and gets its minimal value $-\sigma $ when when $Tk^{2}=-k(\omega_{m}t-T)$. Therefore the mirror state achieving the maximal positive amplification is $\frac{1}{\sqrt{2}}(|0\rangle _{m}+|1\rangle _{m})$ and the state achieving the maximal negative amplification is $\frac{1}{\sqrt{2}}(|0\rangle
_{m}-|1\rangle _{m})$. The two states are nonclassical states since their Wigner functions \cite{Scully97} have a negative part. Fig. 3 shows the Wigner function of the state achieving the maximal negative amplification. It is obvious that the key to understand the amplification is the superposition of the vacuum state and one phonon state of the mirror, which
is due to the retaining of the \emph{Kerr phase}. The amplifications achieved here are somewhat different from the usual weak measurement results \cite{Aharonov88,Pang12} since we choose the post-selected state to be orthogonal to the initial state.

\begin{figure}[tbp]
\includegraphics[width=1\columnwidth]{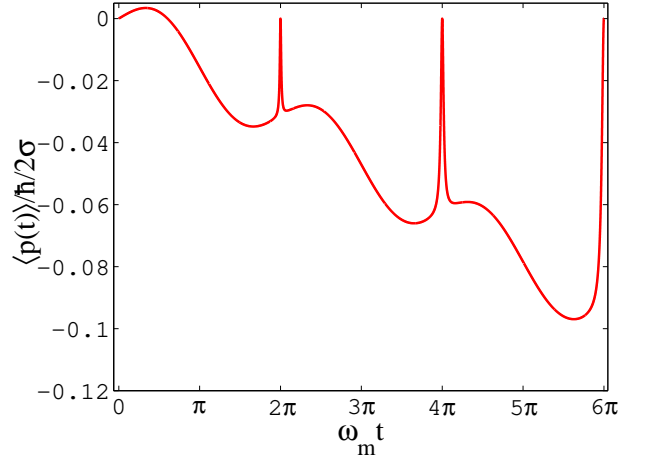}
\caption{The average displacement $\langle p(t)\rangle /\frac{\hbar}{2\sigma}$ of the mirror as a function of $\omega_{m}t$ with $k=0.005$.}
\end{figure}

Next we would like to discuss the amplification of momentum variable $p$ of the mirror. Because the damping coefficient ($\gamma =5\times 10^{-7}$ in \cite{Bouwmeester12}) is very small so we take no account of the damping of the mirror. Substituting Eq. (15) into Eq. (13) which uses $p$ $=-i(c-c^{\dagger })\frac{\hbar }{2\sigma }$ instead of $q$, then we have
\begin{eqnarray}
\langle p(t)\rangle  &=&-i\frac{\hbar }{2\sigma }[\varphi (t)-\varphi ^{\ast}(t)
-e^{-\frac{|\varphi (t)|^{2}}{2}}(e^{i\phi (t)}\varphi (t)  \notag \\
&-&e^{-i\phi (t)}\varphi ^{\ast }(t))]/[2-e^{-\frac{|\varphi (t)|^{2}}{2}}(e^{i\phi (t)}  \notag \\
&+&e^{-i\phi (t)})].
\end{eqnarray}
The average displacement $\langle p(t)\rangle /\frac{\hbar }{2\sigma }$ of the mirror is shown in Fig. 4 as a function of $\omega _{m}t$ with $k=0.005$. It can be seen clearly that the amplification of the mirror momentum $p$ is suppressed around the vibration periods of the mirror $\omega _{m}t=2n\pi$ $(n=1,2,3,\cdots )$.

\section{IV. AMPLIFICATION WITH A PHASE SHIFTER}

\begin{figure}[tbp]
\includegraphics[width=1\columnwidth]{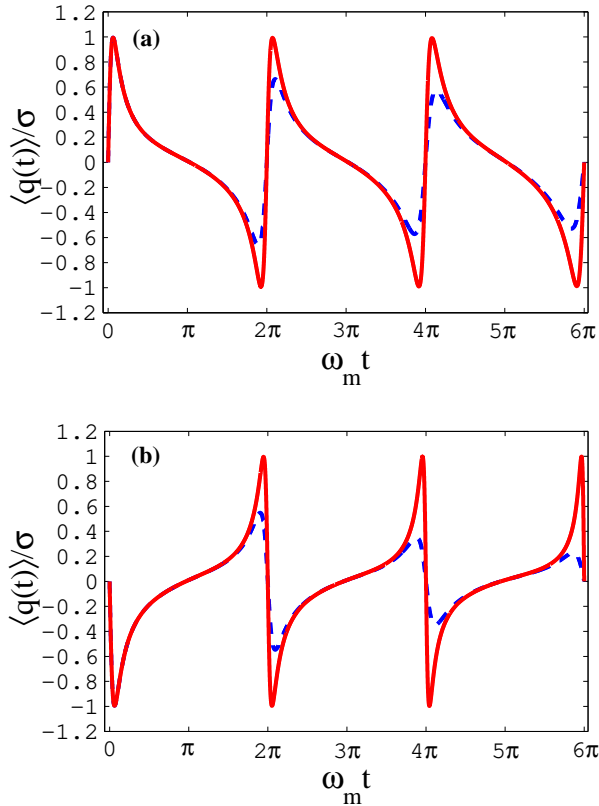}
\caption{The average displacement $\langle q(t)\rangle/\sigma$ for $\theta=0.001$ (a) and $\theta =-0.001$ (b), other parameters are the same as before, i.e., $k=0.005$, $\gamma =0$ (solid line) and $\gamma=0.005$ (dashed line).}
\end{figure}

The key ingredient for amplification is the relative phase between the coherent state and the vacuum state, so we can further add a phase shifter $\theta $ into the interferometer to modulate the relative phase which has been used in \cite{Howell09}. When the phase shifter is added, the initial state of the photon before interacting with the mirror becomes
\begin{equation}
|\psi _{i}(\theta )\rangle =\frac{1}{\sqrt{2}}(e^{i\theta }|1\rangle_{A}|0\rangle _{B}+|0\rangle _{A}|1\rangle _{B}),
\end{equation}
where $\theta $ is positive if the phase shifter is placed in arm $A$ and negative if it is placed in arm $B$. Similar to the previous section, when a photon is detected in the dark port, the expression for the average displacement $\langle q(t)\rangle $ of the mirror is similar to Eq. (14), just with $\phi _{total}(t)=\theta +\phi (t)$ instead of the relative phase $\phi (t)$. Fig. 5(a) and 5(b) show $\langle q(t)\rangle /\sigma $ for $\theta =0.001$ and $\theta =-0.001$, respectively, where other parameters are the same as before, i.e., $k=0.005$, $\gamma =0$ (solid line) and $\gamma =0.005$ (dashed line). In sharp contrast to Fig. 2 the amplification effect can occur around $\omega _{m}t=0$, which means the amplification effect occurs in a short evolution time, here 5(a) is a positive amplification and 5(b) is the negative one. We can understand the phenomenon through a small quantity expansion about time $T$ till the second order
\begin{equation}
\Psi _{os}(t)\approx i(\theta +k^{2}T)|0\rangle _{m}+ik(\omega_{m}t-T)|1\rangle _{m},
\end{equation}
which is similar to Eq. (16). It is very easy to see that even when $T=0$ (or $n=0$) there exists the superposition of the vacuum state $|0\rangle _{m}$ and the one phonon state $|1\rangle _{m}$ since an appropriate phase $\theta $ exists. The reason for the amplification phenomenon around $\omega_{m}t=0$ in Fig. 5 is that there exists the phase shifter offering an enough large relative phase between the coherent state $|\varphi (t)\rangle _{m}$ and the ground state $|0\rangle _{m}$. \textbf{By the comparison of Eq. (16) and Eq. (20), it can be seen that the \emph{Kerr phase} leads to the amplification phenomenon when there is no phase shifter, however, when there is a phase shifter, it is the the total phase (the phase shifter and the \emph{Kerr phase}) that leads to the amplification phenomenon. In Fig. 5} the maximal and the minimal values of $\langle q(t)\rangle /\sigma $ appear when $\omega _{m}t=(1\pm k)T\pm \frac{\theta }{k}$, respectively. The widths of the amplification zones around $T$ in Fig. 5 are also increased compared to Fig. 2. Furthermore we find that around time $\omega _{m}t=2n\pi $ $(n=0,1,2,\cdots )$ the amplifications occur only when $\theta \ll k-k^{2}T$. Thus the amplification effect of optomechanical system can be easily modulated by a phase shifter.

\section{V. DISCUSSION}

In fact, we show that not only the one phonon state can be generated in the orthogonal postselection weak-coupling optomechanics scheme by using a single photon, but also a superposition of the ground state and the one phonon state can be generated if we retain the \emph{Kerr phase} in the model proposed by Pepper et al. When the amplitudes of the ground state and the one phonon state are equal, the maximal amplification effect can occur. When
the coupling strength is not so small or a phase shifter is placed in the scheme, the amplification effect can become prominent.

It is difficult to observe the amplification effect due to the very small post-selected probability of success and the very limited time zones for the appearance of the maximal displaced state. Simultaneously measuring the position and momentum of the mirror is impossible due to quantum uncertainty principle. However it is possible to measure only one quadrature component of the mechanical motion, such as the position, to an arbitrary precision
\cite{Jocobs08}. This idea can be used for a full reconstruction of the mechanical quantum state, extracting its Wigner density using quantum state tomography \cite{Aspelmeyera11}. Another method is to use a qubit to interact with the mirror and offer the nonclassicality of the mirror \cite{Agarwal12}. A determinative generation of different states can use a different hybrid atom-optomechanical setup like in Ref. \cite{Vedral13}. Furthermore we find an outstanding amplification effect with a coherent pointer and the amplification effect can be observed.

\section{VI CONCLUSION}

In our scheme even if the coupling is weak, the displacement of the mirror's position can be amplified in \emph{orthogonal} postselection weak measurement, in contrast with the usual weak measurement regime \cite{Aharonov88,Pang12} which indicates that no amplification effects can be obtained when the post-selected state and the initial state of the system is orthogonal. The amplification effect come from the superpositions of the ground state and the one phonon state of the mirror. The maximal positive and negative amplification states are the equal superpositions of the ground state and the one phonon state. The superposition is due to the non-neglecting of the \emph{Kerr phase} in the paper by Pepper et al. In addition, the amplification effect can be easily modulated by a phase shifter. The results deepen our understanding of the weak measurement, provide us a new method to discuss the weak interaction and the decoherence of the macroscopic superpositions.

Finally, we would like to say this paper has two aspects of meanings. The first one is that this approach provides a simple way to generate nonclassical state of the mirror, so we can investigate some quantum nature of a macroscopic object, such as entanglement and decoherence that have been emphasized in \cite{Bouwmeester12}. The second one is that the optomechanical system offers an ideal platform to investigate the post-selected weak measurement because the motion (displacement) of the mirror is related to an exact dynamics evolution state of the optomechanical system which is suitable for understanding the long-time evolution behaviors.

\section{ACKNOWLEDGMENT}

We thank Ming-Yong Ye, Chang-Shui Yu, Jiong Cheng and Shao-Xiong Wu for useful discussions and suggestions. This work was supported by the National Natural Science Foundation of China under grants No. 11175033.

\end{document}